\documentclass[preprint,aps,amsmath,amssymb,superscriptaddress]{revtex4}

\usepackage{graphicx}
\usepackage{dcolumn}
\usepackage{bm}
   
\begin{document}


\title{Quantum Measurement of a Coupled Nanomechanical Resonator\textemdash Cooper-Pair Box System}

\author{E. K. Irish}
 \affiliation{Department of Physics and Astronomy, University of Rochester}

\affiliation{Laboratory for Physical Sciences\\
8050 Greenmead Drive, College Park MD  20740 USA}
\author{K. Schwab}
 \homepage{http://www.lps.umd.edu}
 \email{schwab@lps.umd.edu}
\affiliation{Laboratory for Physical Sciences\\
8050 Greenmead Drive, College Park MD  20740 USA}

\date{\today}

\begin{abstract}
We show two effects as a result of considering the second-order correction to the spectrum of a nanomechanical resonator electrostatically coupled to a Cooper-pair box. The spectrum of the Cooper-pair box is modified in a way which depends on the Fock state of the resonator.  Similarly, the frequency of the resonator becomes dependent upon the state of the Cooper-pair box.  We consider whether these frequency shifts could be utilized to prepare the nanomechanical resonator in a Fock state, to perform a quantum non-demolition measurement of the resonator Fock state, and to distinguish the phase states of the Cooper-pair box.  
\end{abstract}

\pacs{85.25.Cp, 85.35.Gv, 85.85.+j, 03.65.Ta}
\maketitle

\section{\label{sec:intro}Introduction}
The quantum nature of a mechanical device has yet to be demonstrated.  Manifestations of purely non-classical behavior in a linear resonator include energy quantization and the appearance of Fock states; quantum-limited position-momentum uncertainty; and superposition and entangled states.   Nanomechanical resonators (NR), because of their high frequency  \cite{Huang:2002} ($10~\text{MHz}-1~\text{GHz}$), minute mass ($10^{-15}-10^{-16}~\text{kg}$), and low dissipation ($Q\approx 10^3-10^5$), are expected to be physical systems capable of this behavior under realizable laboratory conditions  \cite{Carr:2001b,Schwab:2001c}.  Coupling single-electron devices to these mechanical systems is expected to provide a realistic means to achieve the standard quantum limit for linear position measurement  \cite{Braginsky:quantum,Zhang:2001,Knobel:2002}, illuminate the transition between quantum and classical behavior  \cite{Polkinghorne:2001,Mozyrsky:2002}, and lead to the generation of squeezed  \cite{Blencowe:1999} and entangled states  \cite{Armour:2002b}.

A fundamental challenge is to observe Fock or number states, the energy eigenstates characteristic of a quantized simple harmonic oscillator.  Techniques to  generate and detect these non-classical states have been elusive; the highly linear nature of the NR at low amplitude, together with linear coupling to the thermal environment through the position coordinate, produces coherent states which are difficult to distinguish from the classical harmonic oscillator.  Additionally, no scheme with sufficient sensitivity and appropriate non-linear coupling to directly detect the Fock states of a NR has yet been proposed and shown to be viable.

In this paper, we show that linear coupling of a NR to a Cooper-pair box (CPB) produces two interesting non-classical effects.  First, the energy levels of the CPB are shifted by the interaction with the NR.  This shift is dependent on the Fock state of the NR.  We will explore the possibility of using spectroscopic measurement of the CPB transition frequency to project a NR into a desired Fock state, and to perform a quantum non-demolition (QND) measurement of the NR Fock state.

Secondly, we show that the resonant frequency of the NR is dependent upon the quantum state of the CPB.  This frequency shift is largest when the CPB is biased to the degeneracy point.  At this point, the eigenstates are two orthogonal equal superpositions of charge, differing only by a phase.  Thus spectroscopy on the NR might be used to distinguish between two states which are indistinguishable by any charge detector  \cite{Devoret:2000}.

\begin{figure}
\includegraphics{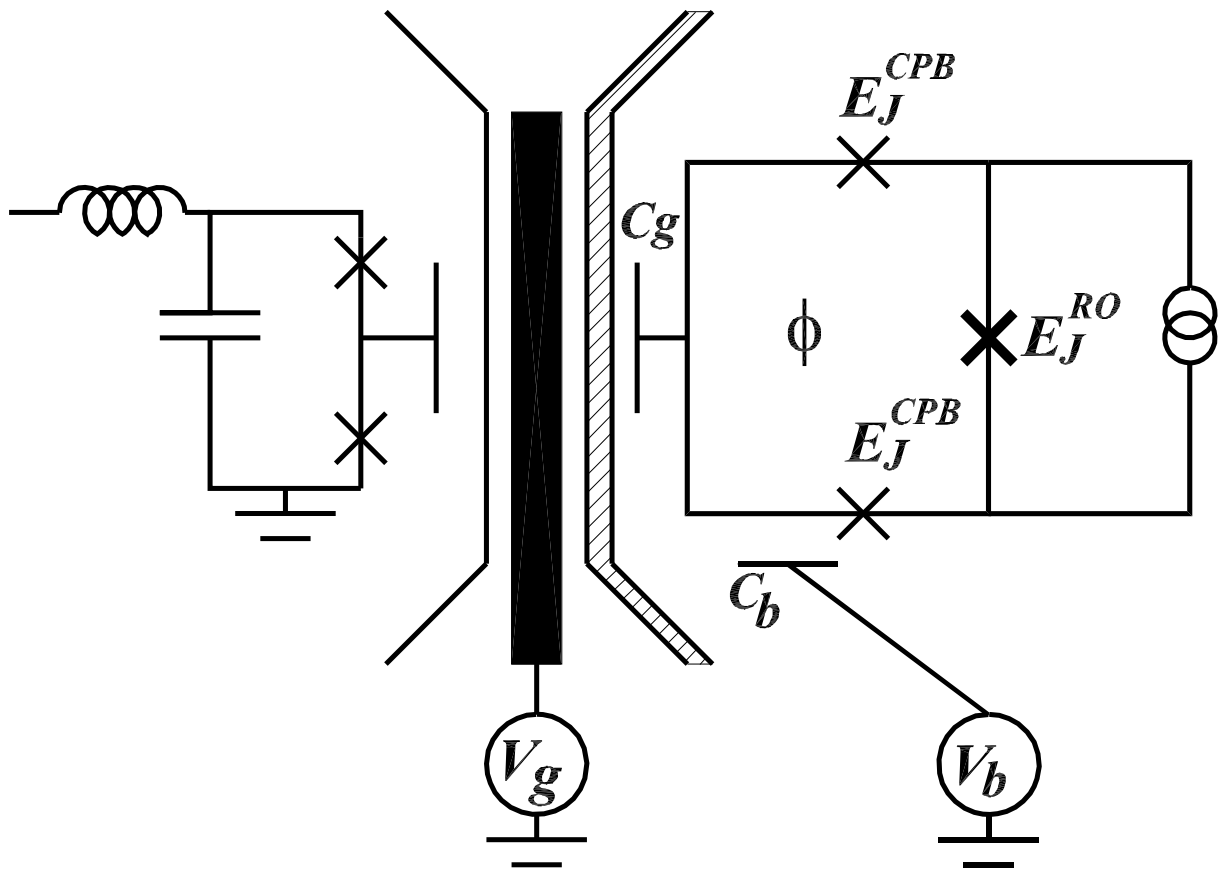}
\caption{\label{fig:schematic} Schematic of coupled CPB-NR system: NR biased with voltage $V_g$ and capacitance $C_g$ to the CPB.  \textit{rf} SET is shown on left to detect the NR position.  The CPB is formed by two junctions with Josephson energy $E_{J}^{CPB}$ biased with flux $\phi$. Read-out of the CPB is accomplished with a large junction, $E_{J}^{RO}$, and current source shown on right. Excitation of the CPB is accomplished by $V_b$ and $C_b$.}
\end{figure}

These effects are both enabled and given relevance by the dramatic experimental results with the CPB  \cite{Vion:2002}.  Vion \textit{et al.} have demonstrated that by biasing a CPB near its degeneracy point and using a pulsed measurement scheme, decoherence times, $\tau_D$, of $500 ~\text{ns}$ are achievable, much longer than $\tau_D \approx 5 ~\text{ns}$ for the bare charge states  \cite{Nakamura:2002}.  A read-out mechanism sensitive to the energy eigenstates rather than the charge states was accomplished using an additional tunnel-junction and high speed current pulses.  Other experimental techniques to distinguish these decoherence-resistant states, such as the method described here, could be very useful.  

In addition, Vion \textit{et al.} have performed high resolution CPB spectroscopy.  Because of the long excited state lifetime, $T_1=2 ~\mu \text{s}$, energy level spectroscopy with resolution of about $10 ~\text{ppm}$ was achieved  \cite{Vion:2002}. Furthermore, Yang \textit{et al.}  \cite{Yang:2002} have achieved $4 ~\text{ppm}$ resolution of the resonant frequency of a $100 ~\text{MHz}$ NR with a $1 ~\text{s}$ measurement time.  Thus subtle frequency shifts of the CPB and the NR which result from coupling may be probed sensitively via spectroscopy.

The implications of these effects are wide reaching. Experimental verification would provide the first evidence that the energy of a nanomechanical system is in fact quantized, and that a mechanical oscillator can be prepared in a number state.  Other closely related systems (two-state system coupled to resonator) such as in mechanical detection of single spins \cite{Sidles:1995,Berman:2000} should be expected to show similar effects. On most general ground, it is hoped that experiments to confirm these predictions will shed light on the nature of the apparent boundary between the classical and the quantum world: is there a limit to the size of an object that can display quantum behavior  \cite{Leggett:2002}? Can we understand the decoherence of ever larger systems?

\section{Energy Shift due to Interaction}
We begin with the Hamiltonian approximating the coupled system shown schematically in Figure \ref{fig:schematic}, where the coupling is given by the electrostatic force between the NR and the CPB  \cite{Armour:2002b}.  We model the NR as a single, simple harmonic mode with resonant frequency $\omega_0$.  As we will show, the largest effects are near the CPB degeneracy point, where two of the charge levels are nearly degenerate. We follow the usual notation as in Ref.  \onlinecite{Makhlin:2001,Armour:2002b} with a few changes for clarity: 

\begin{align*}
H_{TOTAL}&=H_{CPB}+H_{NR}+H_{INT}\\
H_{CPB}&=4E_C(n_g-n-\frac{1}{2})\hat{\sigma}_z-\frac{E_J}{2}\hat{\sigma}_x\\
H_{NR}&=\hbar\omega_0 \hat{a}^\dagger \hat{a} \\
H_{INT}&=\lambda (\hat{a}^\dagger + \hat{a})\hat{\sigma}_z
\end{align*}

where $\hat{a}^\dagger,\hat{a}$ are raising and lowering operators which act only on the NR; $\hat{\sigma}_z, \hat{\sigma}_x$ are Pauli spin matrices operating on the CPB; $n$ is an integer which labels the charge states of the CPB; $n_g=(C_bV_b+C_gV_g)/2e$ where $C_b$ and $V_b$ are the CPB biasing capacitance and voltage and $C_g$ and $V_g$ are the capacitance and voltage between the NR and the CPB; $E_C$ and $E_J$ are the Coulomb and Josephson energies; $\omega_0$ is the unperturbed mechanical frequency; and  $\lambda=-4E_C n_g^{NR} \Delta x_{ZP}/d$ where $n_g^{NR}=C_gV_g/2e$, $\Delta x_{ZP}=\sqrt{\hbar/2m\omega_0}$, which is the zero-point uncertainty of the NR ground state, and $d$ is the distance between the NR and the CPB.

We assume that the Josephson energy of the large read-out junction is much larger than that of the CPB,  $E_{J}^{RO}\gg E_{J}^{CPB}$  \cite{Cottet:2002}.  Because of this, we can approximate the Josephson energy as $E_J=2E_J^{CPB}\cos (\pi\phi/ \phi_0)$ where $\phi$ is the magnetic flux applied to the box and $\phi_0=h/2e$ is the flux quantum. Furthermore, we have not included a term in the Hamiltonian to model the environment since the CPB decoherence time, $\tau_D$, has been measured to be $500 ~\text{ns}$  \cite{Vion:2002}, and the NR is expected to show decoherence times of $1 ~\mu \text{s}$ or longer at temperatures near $10 ~\text{mK}$  \cite{Schwab:2001c,Zurek:1993}. The effects and measurement strategies proposed here do not require coherence on microsecond or longer time scales.

\begin{figure}
\includegraphics*{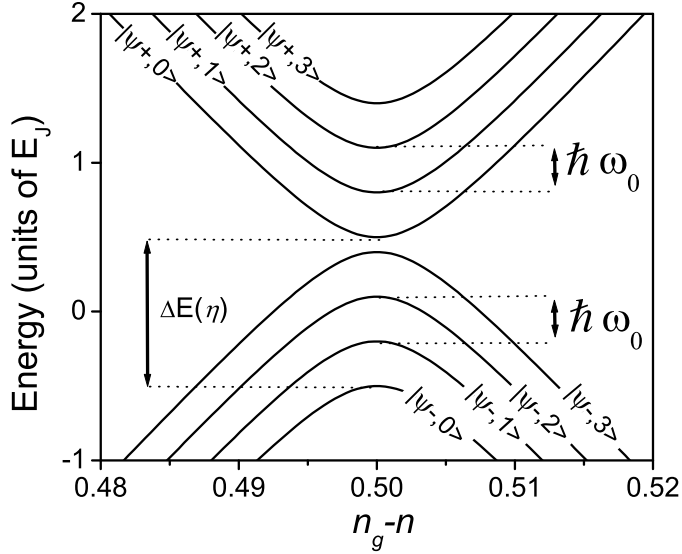}
\caption{\label{fig:unpert} Manifold of unperturbed energy levels of coupled CPB + NR system vs. CPB gate voltage $n_g-n$, near the CPB degeneracy point. The $\psi_{\pm}$ indicates the upper/lower CPB state and the number indicates the NR number state (only lowest four number states are shown). We have assumed $E_C=25E_J$ and $\hbar\omega_0=0.31E_J$. Transitions defining the mechanical frequency $\hbar\omega_0$ and the CPB transition $\Delta E(\eta)$ are shown with arrows.}
\end{figure}

The unperturbed energy levels are given simply by
\begin{align*}
(H_{CPB}+H_{NR})\lvert \psi_{\pm},N\rangle&=E_{\pm,N}^{(0)}\lvert \psi_{ \pm},N\rangle\\
&=(\pm \frac{1}{2} \Delta E(\eta)+N\hbar\omega_0)\lvert \psi_{\pm},N\rangle
\end{align*}
where $N$ is an integer corresponding to the number state of the NR; the unperturbed CPB energy is given by $\Delta E(\eta)=\sqrt{[4E_C(2n+1-2n_g)]^2+E_J^2}$; and the eigenstates expressed in the charge basis are given by
$\lvert \psi_- \rangle = \cos{(\eta/2)}\lvert n \rangle +\sin{(\eta/2)}\lvert n+1 \rangle$ and $
\lvert \psi_+ \rangle = -\sin{(\eta/2)}\lvert n \rangle +\cos{(\eta/2)}\lvert n+1 \rangle$
where $\tan{\eta}=E_J/4E_C(2n+1-2n_g)$.  Figure \ref{fig:unpert} shows the manifold of unperturbed levels as a function of $n_g-n$.

Treating the interaction piece of the Hamiltonian as a perturbation, we calculate the correction to the energy levels to second order: 
\begin{equation}
E_{\pm,N}^{(2)}=E_{\pm,N}^{(0)}+\Delta_{\pm,N}^{(1)}+\Delta_{\pm,N}^{(2)}
\end{equation}
where 
\begin{equation}
\Delta_{\pm,N}^{(1)}= \langle \psi_{\pm},N\rvert H_{INT}\lvert \psi_{\pm},N\rangle=0
\end{equation}
since $\langle N\rvert(\hat{a}^\dagger + \hat{a})\lvert N\rangle=0$, and

\begin{align}
\Delta_{\pm,N}^{(2)}&=\sideset{}{}\sum_{\substack{i,M\neq\pm, N}} \frac{\lvert\langle \psi_i,M\rvert H_{INT}\lvert \psi_{\pm},N\rangle\rvert^2}{E_{\pm,N}^{(0)}-E_{i,M}^{(0)}}\nonumber\\
\label{shift}
&=\lvert \lambda \rvert ^2\Biggl[-\frac{\cos^2{\eta}}{\hbar\omega_0} + \sin^2{\eta}\Bigl[\frac{\pm (2N+1) \Delta E(\eta)+\hbar\omega_0}{\Delta E(\eta)^2-(\hbar\omega_0)^2}\Bigr]\Biggr].
\end{align}

\begin{figure}
\includegraphics{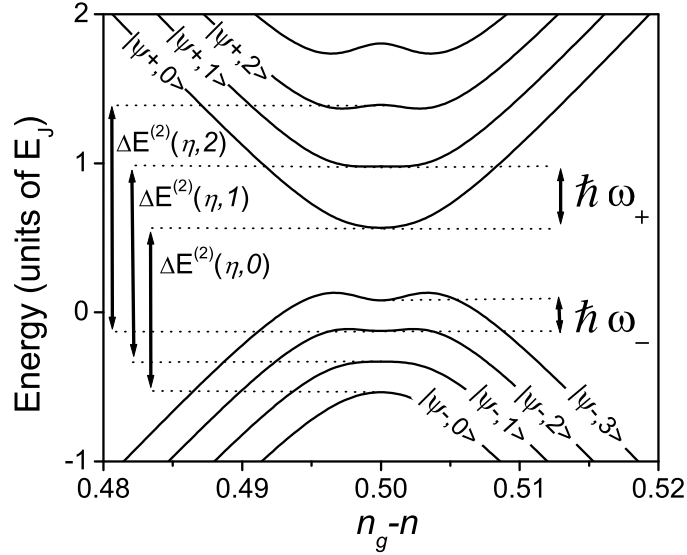}
\caption{\label{fig:spectrum} Manifold of perturbed energy levels of coupled CPB + NR system vs. CPB gate voltage $n_g-n$, near the CPB degeneracy point.  $E_C$ and $E_J$ are as in Figure \ref{fig:unpert}. We have chosen a large value of $\lambda=0.7 \hbar\omega_0$ for illustration purposes only; more realistic values will be given in later figures. Transitions defining the mechanical frequencies $\hbar\omega_{-}$ and $\hbar\omega_{+}$ and the first three CPB transitions $\Delta E^{(2)}(\eta,N)$ are shown with arrows.}
\end{figure}

The perturbed spectrum is shown in Figure \ref{fig:spectrum}.  This simple calculation is the basis of the effects and measurement strategies described in this paper.  This result was shown in Ref.  \onlinecite{Armour:2002a} where the emphasis was on a Lamb shift effect on the CPB from the presence of the zero-point uncertainty of the NR ground state.  In light of recent progress with CPB spectroscopy  \cite{Vion:2002}, this Lamb shift should be observable and would provide evidence for the zero-point motion of a mechanical system.

\begin{figure}
\includegraphics{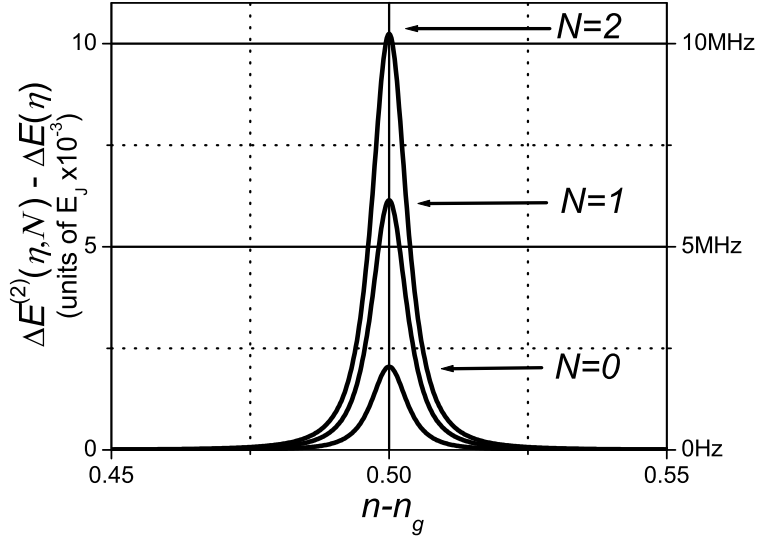}
\caption{\label{fig:CPBshift} Shift in the CPB excitation energy, $\Delta E^{(2)}(\eta,N)-\Delta E(\eta)$, expressed in units of $E_J$ versus CPB gate charge. Values are plotted for the three lowest resonator states.  Mechanical Lamb shift is labeled $N=0$. The right hand scale gives actual values of $(\Delta E^{(2)}(\eta,N)-\Delta E(\eta))/ h$ for experimentally achievable parameters: $E_C=100 ~\mu \text{eV}$, $E_J=4 ~\mu \text{eV}$, $\hbar\omega_0=1.2 ~\mu \text{eV} (300 ~\text{MHz})$, and $\lambda=0.12 ~\mu \text{eV}=0.10\hbar\omega_0$. }
\end{figure}

Up to first order in the perturbation parameter $\lambda$, the new eigenstates are given by:
\begin{align*}
\lvert \psi_{\pm},N\rangle^{(1)}&=\lvert \psi_{\pm},N\rangle+\sum_{\substack{i,M\neq\pm, N}}\lvert \psi_{i},M\rangle\frac{\langle \psi_{i},M\rvert H_{INT}\lvert \psi_{\pm},N\rangle}{E_{\pm,N}^{(0)}-E_{i,M}^{(0)}}\\
&=\lvert \psi_{\pm},N\rangle+\lambda\Biggl[\pm\frac{\cos{\eta}}{\hbar\omega_0}\Big(\sqrt{N+1}\lvert\psi_{\pm},N+1\rangle-\sqrt{N}\lvert\psi_{\pm},N-1\rangle\Big)\\
&-\sin{\eta}\Big(\frac{\sqrt{N+1}}{\pm\Delta E(\eta)-\hbar\omega_0}\lvert\psi_{\mp},N+1\rangle+\frac{\sqrt{N}}{\pm\Delta E(\eta)+\hbar\omega_0}\lvert\psi_{\mp},N-1\rangle\Big)\Biggr]
\end{align*}
We usually wish to bias near the degeneracy point where $\eta=\pi/2$. At this point, the mixing of the new eigenstate $\lvert\psi_{\pm},N\rangle^{(1)}$ is primarily from the nearby mechanical states $\lvert\psi_{\mp},N+1\rangle$ and $\lvert\psi_{\mp},N-1\rangle$.  For reasonable values of  $\lambda$ and low number states this mixing is rather minor.  For example, assuming the ratios $\hbar \omega_0/E_J$ and $\lambda/E_J$ shown in Figure \ref{fig:CPBshift}, the state $\lvert\psi_{+},0\rangle^{(1)}$ includes a contribution from the unperturbed state $\lvert\psi_{-},1\rangle$ with an amplitude of $-0.05$. Also, as will be shown below, the basic structure of the eigenvalues is not changed by this perturbation; the NR states associated with each of the two CPB states remain equally spaced. Because of this, we will drop the superscript on the new eigenstates.

First we consider the effect of the NR on the CPB levels.  Using Eq.~\ref{shift}, we can calculate the energy difference between $\lvert \psi_+,N\rangle$ and $\lvert\psi_-,N\rangle$:  
\begin{align}
\Delta E^{(2)}(\eta,N)&=E_{+,N}^{(2)}-E_{-,N}^{(2)}\nonumber\\
&=\Delta E(\eta) \Bigl[ 1+2\lvert \lambda \rvert ^2 \frac{\sin^2\eta (2N+1)}{\Delta E(\eta)^2 - (\hbar \omega_0)^2} \Bigr].
\end{align}
This transition is shown in Figure \ref{fig:spectrum}.  It is apparent that the CPB energy difference is linearly dependent on the NR number state.  Figure \ref{fig:CPBshift} shows this frequency shift versus $n_g-n$ for the lowest three resonator states and an achievable set of device parameters.  Away from the degeneracy point the only effect of the interaction is to shift the entire structure of energy levels by $-\lvert \lambda \rvert ^2\cos^2{\eta}/\hbar\omega_0$, which is equivalent to altering the zero-point of energy.  As $(n_g-n) \to 1/2$, the state-dependent energy shifts, which are of interest here, emerge. 

Most interestingly, this effect can be used both to monitor and to prepare the NR number states, and can be accomplished as follows.  Suppose the CPB is prepared in the ground state and the mechanical system is in an arbitrary state described by a density matrix in the Fock basis:
\begin{equation}
\hat{\rho}_{initial}=\sum_{N,M=0}^\infty \rho_{N,M} \lvert \psi_{-},N \rangle \langle \psi_{-},M \rvert .
\end{equation}

A $\pi$ pulse is applied to the CPB, where in this case the microwave excitation is tuned to the transition frequency $\Delta E^{(2)}(\eta,J)$, targeting the mechanical state $\lvert J\rangle $. This operation is described by a unitary matrix

\begin{align}
\hat{U} =~ &\lvert \psi_{+}, J \rangle \langle \psi_{-}, J \rvert + \sum_{I \neq J}^{\infty}\lvert \psi_{-}, I \rangle \langle \psi_{-}, I \rvert \nonumber \\
&+ \lvert \psi_{-}, J \rangle \langle \psi_{+}, J \rvert + \sum_{I \neq J}^{\infty} \lvert \psi_{+}, I \rangle \langle \psi_{+}, I \rvert .
\end{align}

The action of $\hat{U}$ on $\hat{\rho}_{initial}$ gives a new density matrix of the coupled system

\begin{align}
\hat{\rho} =~ &\hat{U} \hat{\rho}_{initial} \hat{U}^\dagger\nonumber\\
=~ &\rho_{J,J}\lvert\psi_{+}, J\rangle \langle \psi_{+}, J \rvert 
   + \sum_{N,M\neq J}^\infty \rho_{N,M}\lvert \psi_{-}, N \rangle \langle \psi_{-},M \rvert \nonumber\\
   &+ \sum_{N \neq J}^\infty \rho_{N,J} \lvert \psi_{-}, N \rangle \langle \psi_{+}, J \rvert 
   + \sum_{N \neq J}^\infty \rho_{J,N} \lvert \psi_{+}, J \rangle \langle \psi_{-}, N \rvert .
\end{align}

Next, a current pulse is used to interrogate the state of the CPB, as was done by Vion, \textit{et al.}  Ideally, this current pulse may be described by projective measurement operators $\hat{M}_{-} = \lvert \psi_{-} \rangle \langle \psi_{-} \rvert \otimes I_{NR} \nonumber$ and $\hat{M}_{+} = \lvert \psi_{+} \rangle \langle \psi_{+} \rvert \otimes I_{NR}$, where $\hat{M}_{m}$ corresponds to measuring the CPB in state $\lvert \psi_{m} \rangle$, leaving the NR state unaffected. The final density matrix resulting from a measurement which gives the result $m$ is given by

\begin{equation}
\hat{\rho}_{m} = \frac{\hat{M}_{m} \hat{\rho} \hat{M}_{m}^{\dagger}} {Tr(\hat{M}_{m}^{\dagger} \hat{M}_{m} \hat{\rho})} \nonumber.
\end{equation}

Applying this to the $\hat{\rho}$ found above gives two possible final density matrixes

\begin{align}
\hat{\rho}_{-} &= \frac{\sum_{N,M \neq J}^{\infty} \rho_{N,M} \lvert \psi_{-}, N \rangle \langle \psi_{-}, M \rvert}{1-\rho_{J,J}} \\
\hat{\rho}_{+} &= \lvert \psi_{+}, J \rangle \langle \psi_{+}, J \rvert .
\end{align}

Thus with probability $\rho_{JJ}$ this procedure has the effect of both taking an arbitrary system distribution and creating a pure Fock state as well as providing information of this preparation to the experimenter.


An important consideration is the lifetime, and the associated line broadening, of the NR number state. For the photon case, it has been shown that the lifetime of a Fock state $\lvert N \rangle$ interacting with a zero-temperature dissipative reservoir is given by $\tau_N = 1/N\gamma$, where $\gamma$ is the cavity decay rate  \cite{Lu:1989}. Therefore we expect that the lifetime of a Fock state of the NR will be similarly related to the decay rate of the resonator $\gamma_{NR}$.  Assuming typical NR properties of $\omega_0/2\pi=300 ~\text{MHz}$ and $Q=10^4$, we find that $1/\gamma_{NR} \approx 5.3 ~\mu \text{s}$, which corresponds to a linewidth of approximately $30 ~\text{kHz}$\footnote{We have recently measured a NR with frequency of $10MHz$ and a Q=10,000 at 4.2K, rising to Q=170,000 at 20mK.  We expect higher frequency resonators to show similar increase in Q at very low temperatures.}.  At a temperature $T = 20 ~\text{mK}$, the average thermal excitation $n_{th} = (e^{\hbar \omega_0 / k_B T} - 1)^{-1} \approx 0.95$; the thermal equilibrium state is reasonably close to the ground state.  Thus up to Fock state $N\approx 30$ we expect the linewidths to be less than $1 ~\text{MHz}$.  For the CPB alone, the linewidth achieved by Vion \textit{et al.} in Ref.  \onlinecite{Vion:2002} was about $0.8 ~\text{MHz}$.  The maximum separation between peaks corresponding to adjacent $N$ values for parameters given in Figure \ref{fig:CPBshift} is around $4 ~\text{MHz}$; the transitions $\Delta E^{(2)}(\eta,N)$ should be well resolved.  

The energy shift of the CPB may also be a basis for performing another type of QND measurement on the resonator number state, following a close analogy to the procedure demonstrated in cavity quantum electrodynamics (CQED) for performing QND measurement of microwave cavity photons  \cite{Brune:1992,Nogues:1999}.  The procedure relies on Ramsey interferometry  \cite{Ramsey:1950} performed on the CPB  \cite{Vion:2002}.  This is accomplished by beginning with the CPB in the ground state $\lvert \psi_- \rangle$, with a large static coupling $\lambda$ to the NR, and biased away from  degeneracy to a point where the transition frequency is not a function of the NR number state, i.e. $\Delta E(\eta,N+1)-\Delta E(\eta,N)<< \hbar/T_1$.  The state $(\lvert \psi_- \rangle +e^{i\delta_0}\lvert \psi_+ \rangle)/\sqrt{2}$ is prepared by a microwave $\pi /2$ pulse.  The voltage between the NR and CPB is increased adiabatically to the CPB degeneracy point, so that the energies of the CPB states become dependent on the NR number state.  After a time $t$ the prepared state evolves into $(\lvert \psi_- \rangle +e^{i \phi(t,N)}\lvert \psi_+ \rangle)/\sqrt{2}$ where $\phi(t,N)=\Delta E^{(2)}(\eta,N)\cdot t/\hbar+\delta_0$.  The gate voltage is then adiabatically switched back to the initial value, followed by another $\pi/2$ pulse and, finally, measurement of the CPB state.  The probability to find the CPB in the lower state is found to be $P_{-}(t,N)=\big(1-\sin{\phi(t,N)}\big)/2$.  Assuming the parameters shown in Figure \ref{fig:CPBshift} and an interaction time of $t \approx 60 ~\text{ns}$, which is much smaller than all of the relaxation and decoherence times in the system, a substantial phase difference of $\Delta\phi(t)=\phi(t,N+1)-\phi(t,N) \approx \pi/2$ is developed in the CPB state between the $N$ and $N+1$ NR Fock state.  Since the NR state is not destroyed, this sequence can be repeated several times within the lifetime of the NR Fock state to determine $P_{-}(t,N)$ which determines $N$.  The QND aspect of this measurement technique is described in the Appendix.

The spectroscopic method and the Ramsey interferometry method of creating Fock states may be viewed as complementary schemes, in the following sense.  Using spectroscopy, a given number state is targeted, although it may not be created every time.  With the Ramsey method, the oscillator will certainly end up in a number state, but which state will be created is probabilistically determined and not known in advance.

\begin{figure}
\includegraphics{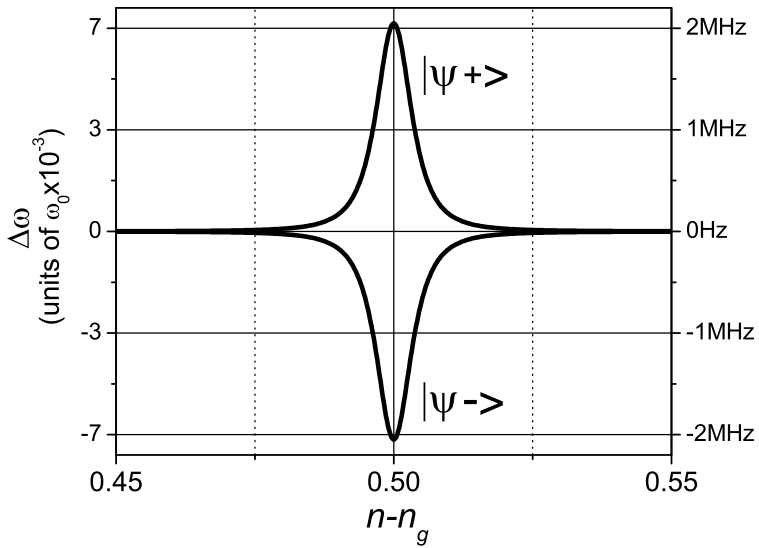}
\caption{\label{fig:NRshift} Shift in the nanomechanical resonator frequency, $\Delta \omega=\omega_{\pm}(\eta)-\omega_0$, expressed in units of $\omega_0$ (frequency) versus CPB gate charge for both the $\lvert \psi_- \rangle$ and $\lvert \psi_+ \rangle$ CPB state.  The right hand scale gives actual values of $\Delta\omega/2\pi$ for the same parameters given in Figure \ref{fig:CPBshift}.}
\end{figure}

Next we consider the effect of the CPB on the NR energy levels.  It is apparent from Eq.~\ref{shift} that the energy levels of the resonator depend upon the  CPB state,  resulting in a shifted mechanical frequency:
\begin{equation}
\hbar\omega_\pm(\eta)=\hbar\omega_0\pm 2\lvert \lambda \rvert ^2\frac{\sin ^2 \eta \Delta E(\eta)}{\Delta E(\eta)^2-(\hbar \omega_0)^2} ,
\end{equation}
where $\pm$ corresponds to the state $\lvert \psi_{\pm} \rangle$ of the CPB.
  Notice that to second order, the mechanical resonator remains linear: the energy levels are equally spaced. Figure \ref{fig:NRshift} shows this shift for the same parameters as used in Figure \ref{fig:CPBshift}. 

The parameters used in Figure \ref{fig:NRshift} should be experimentally achievable, but may be challenging to reach.  However, this effect should be apparent even for lower frequency resonators with rather weak coupling to the CPB.  For instance, the maximum frequency shift is $\Delta\omega =130 ~\text{Hz}$ for a $50 ~\text{MHz}$ resonator with a coupling of $\lambda=0.005\hbar\omega$ and CPB parameters as in Figure \ref{fig:CPBshift}.  This is much larger than the frequency resolution which has been achieved with a $100 ~\text{MHz}$ NR \cite{Yang:2002}, and can be observed by simply measuring the resonant frequency while slowly sweeping the CPB gate bias.

The shift of the mechanical frequency can be used to read-out the state of the CPB.  After the CPB has been prepared in the desired state, one can send a sudden electrostatic drive to the resonator in a time which is short compared to the CPB energy relaxation time, $T_1=2 ~\mu \text{s}$ \cite{Vion:2002}.  The frequency of this drive is chosen to be either $\omega_+$ or $\omega_-$, which excites the NR if the CPB is in the corresponding state $\lvert \psi_+\rangle$ or $\lvert \psi_-\rangle$.  After this sudden drive the response of the mechanical system can be measured, where the detection (or absence) of motion would indicate the state of the CPB.  The final measurement of the NR must be accomplished within the energy relaxation time of the NR.  This scheme can be accomplished with a resonator of frequency $100 ~\text{MHz}$ and $Q=10^4$, and a $1 ~\text{mV}$, $200 ~\text{ns}$ pulse applied from a gate with capacitance $20 ~\text{aF}$ and biased with $10 ~\text{V}$.  Such a pulse will drive the NR to an amplitude of $1\cdot 10^{-12} ~\text{m}$ giving a signal-to-noise ratio of about $10$ using a \textit{rf} SET position detector with displacement resolution of $3\cdot 10^{-16} ~\text{m/}\sqrt{\text{Hz}}$ \cite{Zhang:2001}.  This could provide a mechanical means to distinguish the decoherence-resistant and difficult-to-detect phase states of the CPB.

\section{Comments and Conclusion}
It is interesting to point out that these energy shifts disappear if $E_J$ or $\Delta x_{ZP}\rightarrow 0$, i.e. if the quantum nature of the electronic system or the mechanical device is eliminated.  Measurement of these shifts would provide the first evidence for the validity of a quantum description of the center-of-mass coordinate of a macroscopic mechanical device, a device composed of $10^9$ atoms.  Furthermore, these effects offer the first proposal of a viable scheme to detect and prepare non-classical mechanical states, the Fock states.

Detection of the CPB energy shift from the NR ground state, $\Delta E^{(2)}(\eta,0)$, which is analogous to the Lamb shift \cite{Armour:2002a}, would provide proof of mechanical zero-point fluctuations. In light of demonstrated CPB spectroscopy \cite{Vion:2002} and the size of the shift, this effect appears to be measurable.  This would join a very small number of experiments \cite{Lamb:1947,Sparnaay:1958,Brune:1994} which are sensitive to zero-point energy of any kind.

The physics of the Hamiltonian described here is rather general and may apply in other similar systems, such as a NR coupled to a single electron or nuclear spin \cite{Sidles:1995,Berman:2000}, or a CPB coupled to a LC resonator or equivalent circuit \cite{Marquardt:2001,Saidi:2001}.  A connection to CQED may be made by noting that at the charge degeneracy point $(n_g-n=1/2)$ the Hamiltonian given here, rewritten in the basis of non-interacting energy eigenstates, becomes identical to the two-level atom, single cavity mode Hamiltonian of CQED:
\begin{equation*}
H(\eta = \frac{\pi}{2}) \to -\frac{1}{2}E_J\hat{\rho}_z + \hbar \omega_0 \hat{a}^{\dagger} \hat{a} - \lambda \hat{\rho}_x (\hat{a}^\dagger + \hat{a})
\end{equation*}
where $\hat{\rho}_{x} \equiv \cos \eta \hat{\sigma}_{x} - \sin \eta \hat{\sigma}_{z}$ and $\hat{\rho}_{z} \equiv \sin \eta \hat{\sigma}_{x} + \cos \eta \hat{\sigma}_{z}$ are Pauli spin matrices operating in the energy eigenbasis rather than in the charge basis.  For the situation described here, the detuning parameter is very large since $2\Delta E(\pi/2)/\hbar\omega_0\gg 1$.  In this regime, which is not commonly considered in quantum optics, the rotating wave approximation is not valid, so the Hamiltonian does not reduce to the Jaynes-Cummings Hamiltonian.  Nevertheless, similar energy shifts occur in CQED systems and have been observed in experiments \cite{Brune:1994}.   

Furthermore, the NR+CPB system should be able to achieve the strong coupling regime where the Rabi frequency is much greater than both the NR decay rate $\gamma_{NR}$ and the CPB lifetime $1/T_1$: $\lambda/\hbar \gg (\gamma_{NR},1/T_1)$.  An achievable value for this ratio is $(\lambda/\hbar)^2\cdot T_1/\gamma_{NR}\sim 10^5$, assuming $\lambda=0.12~\mu\text{eV}$, $1/\gamma_{NR}=5~\mu \text{s}$, and $T_1=2~\mu \text{s}$. The close analogy to CQED begs for careful examination in order to understand what new parameter space may be explored by mechanical systems coupled to two-state quantum systems. This will be the subject of future work.

If these effects are experimentally achievable, then a wealth of physical phenomena should be possible.  For instance, by realizing energy spectra as shown in Figure \ref{fig:spectrum}, mechanical cooling may be possible by driving the transition sequence $\vert \psi_{-} ,N\rangle \to \vert \psi_{+}, N-1\rangle$, which is then followed by the natural decay of the CPB state to $\vert \psi_{-} ,N-1\rangle$, resulting in the adsorption of one mechanical quantum.  This can be accomplished by applying the appropriate microwave drive to the CPB and is very similar to side-band cooling as is done with atomic ion trap experiments \cite{Wineland:1987,Heinzen:1990}.  

Recently, a $1 ~\text{GHz}$ NR with $Q\sim 500$ has been reported \cite{Huang:2002} which will allow the direct coupling of a mechanical system which is resonant with the CPB energy splitting.  Assuming that the NR and the CPB are resonant at the degeneracy point, the Hamiltonian in this case takes the very familiar Jaynes-Cummings form:
\begin{align*}
H_{INT}&=-\lambda(\hat{a}^\dagger + \hat{a})\hat{\rho}_x=-\lambda(\hat{a}^\dagger + \hat{a})(\hat{\rho}_- + \hat{\rho}_+)\\
&\approx -\lambda(\hat{a}^\dagger\hat{\rho}_- + \hat{a}\hat{\rho}_+)
\end{align*}
where $\hat{\rho}_+$ and $\hat{\rho}_-$ are the CPB raising and lower operators. In the last equation we have used the rotating wave approximation and have dropped the energy non-conserving terms $\hat{a}^\dagger\hat{\rho}_+$ and $\hat{a}\hat{\rho}_-$.  This clearly describes the coherent exchange of a single quantum between the mechanical system and the CPB, at the Rabi frequency $\lambda/h$. 

This is a direct analogy of the situation in CQED and should allow similar phenomena. For instance, cooling the resonator could be accomplished by preparing the CPB in the $\vert \psi_{-}\rangle$ state and biased slightly away from degeneracy, with the coupling to the NR such that the Rabi frequency is smaller than the CPB or NR transition frequency.  By changing the bias adiabatically such that the CPB and the NR are resonant for half the Rabi time, the CPB will be promoted to the excited state at the expense of one mechanical quantum.  This deterministically changes the system state from $\vert \psi_{-},N\rangle\rightarrow\vert \psi_{+},N-1\rangle$ state and removes one quantum from the NR.  After decay of the CPB into $\vert \psi_{-} ,N-1\rangle$, this process could be repeated.  

In addition, resonant coupling of the NR to a two-level quantum system may provide a method to exchange quanta between two-level qubits.  One could use a nanomechanical ``bus'' to couple charge qubits, much in the same way as single atoms are coupled in an ion trap through the quantized vibrational states \cite{Cirac:1995,Brown:2001}, or an atom is coupled resonantly to a electromagnetic cavity\cite{Kimble:1994,Haroche:1994}.  Nanomechanical resonators offer high frequency, high quality factor, and the potential for tight coupling in a very compact object, much smaller than electromagnetic resonators which have been proposed for this purpose  \cite{Makhlin:1999}: a $1 ~\text{GHz}$ mechanical resonator is $ \sim 1 ~\mu \text{m}$ long, while a $1 ~\text{GHz}$ $\lambda/4$ strip-line resonator is  $\sim 2 ~\text{cm}$ long.

In conclusion, we have shown that both the resonant frequency of a nanomechanical resonator and the energy levels of a Cooper-pair box are shifted when the two devices are capacitively coupled.  These shifts are largest at the degeneracy points of the box where the eigenstates are equal superpositions of the two charge states, differing only by a phase.  Experiments to use these effects to manipulate and measure the quantum state of the nanomechanical system and the Cooper-pair box appear viable and are under investigation.  The effects and proposed techniques discussed here further develop the fully quantum treatment of electronic and mechanical devices, a regime we call Quantum Electro-Mechanics.

\begin{acknowledgments}
We would like to acknowledge helpful conversations with Miles Blencowe, Andrew Armour, Nicholas Bigelow, Michael Wulf, Ivar Martin, Carlos Sanchez,  Xuedong Hu, Sankar Das Sarma, Andrew Skinner, and Arthur Vandelay. This work has been supported by the National Security Agency. E.K.I. acknowledges support from a National Physical Sciences Consortium fellowship.
\end{acknowledgments}
\appendix

\section{Quantum Non-Demolition Measurement of Resonator Fock state}
The analysis of the QND aspect of the Ramsey interference technique follows closely the work of Imoto, \textit{et al.} \cite{Imoto:1985}, and Brune, \textit{et al.} \cite{Brune:1992}, and is outlined here as a further illustration of the similarities between the CPB-NR system and atom-cavity systems.  The resonator is the quantum system under study (cavity field), and the CPB is the quantum probe (atom).  The system quantity which we wish to measure is $\hat{A}_S=\hat{a}^\dagger \hat{a}$.  In the Ramsey interference scheme, the last step is to rotate the CPB state by $\pi/2$ and perform a projection onto the eigenbasis.  Thus the measured probe quantity is
\begin{equation}
\hat{A}_P=\frac{\hat{\rho}_{+}-\hat{\rho}_{-}}{2i}
\end{equation}
where $\hat{\rho}_{+}$, $\hat{\rho}_{-}$ are the CPB raising and lowering operators.   

Assuming that the CPB is biased at the degeneracy point and dropping all constant terms, we can write the perturbed energy of the state $\vert\psi_{\pm},N\rangle$ as:
\begin{align}
E^{(2)}_{\pm,N}(\eta = \frac{\pi}{2}) &= \pm \frac{E_J}{2} + N\hbar\omega_{0} \pm \lvert\lambda\rvert^2\frac{(2N+1)E_J}{E_J^2-(\hbar\omega_{0})^2}\\
  &=  \pm \frac{E_J}{2}
\biggl[ 1+\frac{2\lvert\lambda\rvert^2}{E_J^2-(\hbar\omega_{0})^2}\biggr] + N\hbar\omega_{0} \pm \frac{2\lvert\lambda\rvert^2 E_J}{E_J^2-(\hbar\omega_{0})^2}N.
\end{align}
Noting that $N$ is the eigenvalue of $\hat{a}^{\dag} \hat{a}$ and that $\pm 1$ is the eigenvalue of $\hat{\rho}_{z} = (2\hat{\rho}_{+} \hat{\rho}_{-}-\hat{I})$, the effective interaction Hamiltonian is found, after some algebra, to be:
\begin{equation}
H^{(2)}_{INT} = -\frac{2\lvert\lambda\rvert^2 E_J} {E_J^2 - (\hbar\omega_{0})^2}\hat{a}^\dagger \hat{a} \hat{\rho}_{+}\hat{\rho}_{-}
\end{equation}
where $\hat{a}^\dagger$, $\hat{a}$ are the usual NR raising and lowering operators.  This has the same form as the dispersive, Kerr-type effect utilized for QND measurements in quantum optics\cite{Imoto:1985,Levenson:1986}.

It is not difficult to show that this system satisfies the requirements for a QND measurement scheme of the resonator Fock state\cite{Imoto:1985,Brune:1992}.  The first requirement is that $H_{INT}$ is a function of $\hat{A}_S$: $\partial H_{INT}/\partial \hat{A}_S\neq 0$.  Next, the dynamics of $\hat{A}_P$ should depend upon the interaction Hamiltonian, $[\hat{A}_P,H_{INT}]\neq 0$, while the measured quantity should not, $[\hat{A}_S,H_{INT}]= 0$.  Finally, the system Hamiltonian should not be a function of the conjugate variable to the measured system quantity, which is phase: $\partial H_{S}/\partial \hat{\phi}=0$.  It is clear that the system described above does indeed satisfy these requirements.


\end{document}